\documentclass[conference]{IEEEtran}
\IEEEoverridecommandlockouts

\usepackage{cite}
\usepackage{amsmath,amssymb,amsfonts}
\usepackage{algorithmic}
\usepackage{graphicx}
\usepackage{textcomp}
\usepackage{xcolor}

\usepackage{amsmath, calc}
\usepackage{graphicx}
\usepackage{tcolorbox}
\usepackage{multirow}
\usepackage{url}
\usepackage{hyperref}
\usepackage{pgfplots}
\pgfplotsset{compat=1.18}
\usepackage{textcomp}
\usepackage{placeins}
\usepackage{fontawesome}
\usepackage{comment}
\usepackage{tabularx}
\usepackage{enumitem}
\usepackage{booktabs}
\usepackage{comment}
\usepackage{tabu}

\usepackage{hyperref}

\usepackage{framed}

\usepackage{tcolorbox}
\usepackage{fancybox} 

\usepackage{tabularx}

\usepackage{makecell}
\usepackage{multirow}
\usepackage{csquotes}
\usepackage{graphicx}
\usepackage{orcidlink}
\usepackage{multicol}
\usepackage{colortbl}
\usepackage{fancybox,framed}
\usepackage{tikz}

\def\BibTeX{{\rm B\kern-.05em{\sc i\kern-.025em b}\kern-.08em
    T\kern-.1667em\lower.7ex\hbox{E}\kern-.125emX}}
\begin{document}

\makeatletter
\newcommand{\linebreakand}{%
  \end{@IEEEauthorhalign}
  \hfill\mbox{}\par
  \mbox{}\hfill\begin{@IEEEauthorhalign}
}
\makeatother

\newcommand\copyrighttext{%
  \footnotesize \textcopyright~2025 IEEE. This document is a preprint. Personal use of this material is permitted.
  Permission from IEEE must be obtained for all other uses, in any current or future
  media, including reprinting/republishing this material for advertising or promotional
  purposes, creating new collective works, for resale or redistribution to servers or
  lists, or reuse of any copyrighted component of this work in other works. 
  DOI: \href{https://doi.org/10.1109/ICSE-SEIP66354.2025.00063}{10.1109/ICSE-SEIP66354.2025.00063}
}

\newcommand\copyrightnotice{%
  \begin{tikzpicture}[remember picture,overlay]
  \node[anchor=north,yshift=-12pt] at (current page.north) {\fbox{\parbox{\dimexpr\textwidth-\fboxsep-\fboxrule\relax}{\copyrighttext}}};
  \end{tikzpicture}%
}

\title{Automating Explanation Need Management in App Reviews: A Case Study from the Navigation App Industry}

\author{
    \IEEEauthorblockN{
        Martin Obaidi, 
        Hannah Deters, \\
        Jakob Droste, 
        Marc Herrmann, \\
        Kurt Schneider}
    \IEEEauthorblockA{\textit{Leibniz University Hannover} \\
    \textit{Software Engineering Group}\\
    Hannover, Germany \\
    \{firstname.lastname\}@inf.uni-hannover.de}
    \and
    \IEEEauthorblockN{
        Nicolas Voß}
    \IEEEauthorblockA{\textit{Graphmasters GmbH} \\
    Hannover, Germany \\
    nicolas.voss@stud.uni-hannover.de}
    \and
    \IEEEauthorblockN{
        Jannik Fischbach}
    \IEEEauthorblockA{\textit{Netlight Consulting GmbH and fortiss GmbH}\\
    Munich, Germany \\
    jannik.fischbach@netlight.com}
}

\maketitle

\copyrightnotice
\vspace{-2ex}

\begin{abstract}
Providing explanations in response to user reviews is a time-consuming and repetitive task for companies, as many reviews present similar issues requiring nearly identical responses. To improve efficiency, this paper proposes a semi-automated approach to managing explanation needs in user reviews. The approach leverages taxonomy categories to classify reviews and assign them to relevant internal teams or sources for responses. 2,366 app reviews from the Google Play Store and Apple App Store were scraped and analyzed using a word and phrase filtering system to detect explanation needs. The detected needs were categorized and assigned to specific internal teams at the company Graphmasters GmbH, using a hierarchical assignment strategy that prioritizes the most relevant teams. Additionally, external sources, such as existing support articles and past review responses, were integrated to provide comprehensive explanations. The system was evaluated through interviews and surveys with the Graphmasters support team, which consists of four employees. The results showed that the hierarchical assignment method improved the accuracy of team assignments, with correct teams being identified in 79.2\% of cases. However, challenges in interrater agreement and the need for new responses in certain cases, particularly for Apple App Store reviews, were noted. Future work will focus on refining the taxonomy and enhancing the automation process to reduce manual intervention further.
\end{abstract}


\section{Introduction}
\label{sec:intro}

Explainability is a non-functional requirement that can significantly impact software quality~\cite{Köhl2019,Chazette2020}. To effectively integrate explainability into a system, it is essential to identify the actual needs of users~\cite{Droste2024}. To this end, direct feedback from real users, such as app store reviews or other support channels like email or customer support hubs, can be analyzed to understand these needs~\cite{unterbusch2023explanation}. The Google Play Store and Apple App Store represent the largest market share for such feedback~\cite{StoreStatistiksEnglish2024}. Responding to these requests requires time and individual handling, leading to increased operational costs for companies~\cite{Vermeulen2018}. A fully automated approach to handling app store reviews is challenging to implement due to the complexity and the limited dataset available for each app~\cite{Marcus2018}.

Previous work has developed approaches for automatically generating suggested answers from customer reviews~\cite{Kim2022}. However, some of the explanation needs mentioned in reviews may also arise for other users. Therefore, integrating these answers directly into the system as explanations could improve software quality. To achieve this, reviews must be analyzed for explanation needs and, ideally, forwarded directly to the person responsible. That person can then decide whether the request should be integrated into the system as an explanation or merely addressed within the app store.

To further facilitate the analysis of these requests, a taxonomy can be used to categorize user reviews based on system-specific or non-specific elements~\cite{Panichella2015, Droste2024, Tsakalakis2022}. This paper employs the taxonomy of Droste et al.~\cite{Droste2024} to evaluate explanation needs.
The goal of this paper is to streamline and partially automate the process of addressing user explanation needs. App reviews are classified into taxonomy categories, which are then used to assign a relevant reference point (person or team) within the company. The dataset for this taxonomy-to-reference mapping was derived from interviews and a survey conducted with Graphmasters GmbH. Additionally, a source containing the necessary explanation was identified for each review using an API from the Graphmasters support platform and existing responses from the Google Play Store.
Given the variability and specificity of app reviews, a semi-automated process is proposed, with manual verification by a requirements engineer at each step. This semi-automated process was evaluated at Graphmasters GmbH through interviews and surveys, demonstrating its applicability.

The rest of this paper is structured as follows: In Section~\ref{sec:background}, we present related work and background details. The development of the study design is described in Section~\ref{sec:research}. Section~\ref{sec:results} summarizes the results, which are discussed in Section~\ref{sec:discussion}, before concluding the paper in Section~\ref{sec:conclusion}.

\section{Background and Related Work}
\label{sec:background}

This section provides an overview of the background and related research on explainability in software engineering.

\subsection{Background}
\label{sec:erklearbarkeit}
The concept of \textit{explanation needs} originates from requirements engineering, describing a type of information that a user requires to achieve a goal in a particular context. Explainability is a non-functional requirement because it primarily enhances user satisfaction rather than being critical to system functionality. Chazette et al.~\cite{Chazette2021} offer a definition that specifically fits the context of this work:

\begin{quote} \textit{A navigation app is considered explainable with respect to its route guidance relative to a user in the context of road usage, only if there is a support team member who can help the user understand the app's route guidance by providing a support article or an answer in an app store.} \end{quote}

This definition illustrates that explainability is not solely achieved through in-system self-explanations but can also be improved through external explanations, such as support articles~\cite{Chazette2021}.

In modern software systems, explainability significantly impacts customer satisfaction~\cite{deters2024ExplainabilityAndUX}. Explanations that deliver the correct information at the right time positively influence the user experience~\cite{deters2024ExplainabilityAndUX}, enhancing overall satisfaction. In app reviews, user satisfaction is often measured using a star-rating system ranging from one to five stars, where one star indicates poor satisfaction and five stars signify a positive experience. Well-implemented explanations can improve the success of software that is made available on platforms with public review functions.

To support the management of explainability requirements, Droste et al.~\cite{Droste2024} developed a taxonomy that categorizes explanation needs based on the type of information required. They distinguish five main categories. \textit{System Behavior} covers information on how the system operates, especially in cases of unexpected behavior or bugs. \textit{Interaction} explanations aim to help the user achieve a specific goal with the system. \textit{Privacy \& Security} addresses concerns regarding privacy information and security measures implemented in the system. \textit{Domain Knowledge} involves system-specific elements and domain-specific terminology. Lastly, \textit{User Interface} focuses on queries related to design decisions, particularly when changes are made to the user interface.

We used an extended dataset by Droste et al.~\cite{Droste2024}, which included two additional categories: \textit{Meta Information} and \textit{Timing}. \textit{Timing} can be additionally used when the explanation needs concern changes in the past or desired changes in the future, while \textit{Meta Information} describes explanation needs that cannot be easily assigned to the other categories. In this study, we applied the extended taxonomy developed by Droste et al.~\cite{Droste2024} to categorize explanation needs.

\subsection{Related Work}

Several studies have explored the automatic identification and handling of explanation needs in software systems. 
Liu et al.~\cite{Liu2023} examined the performance of ChatGPT compared to traditional recommendation systems, which are typically designed to solve task-specific problems. Despite lacking the fine-tuning that traditional methods offer, ChatGPT outperformed these systems in generating useful explanations for complex tasks.

Unterbusch et al.~\cite{unterbusch2023explanation} investigated various natural language processing (NLP) models to automatically detect explanation needs in app reviews. They found that the BERT model outperformed rule-based approaches, achieving an $F_\beta$\textit{-score} of 0.93 with a weighting of $\beta$ = 19.52. Although rule-based models were generally less effective, they performed comparably to BERT in one specific app. However, both models showed a significant drop in precision, generating a large number of false positives.

Horstmann et al.~\cite{Horstmann2023} analyzed the self-explanation capabilities of AI systems, evaluating explanations from a psychological perspective based on user needs. They also considered the legal and ethical requirements that AI systems must follow, particularly emphasizing the importance of data protection and informational self-determination.

Tsakalakis et al.~\cite{Tsakalakis2022} proposed a taxonomy for classifying explanations. Their taxonomy is divided into nine dimensions aimed at meeting regulatory and business requirements, while also improving the transparency of automatically generated explanations.

Panichella et al.~\cite{Panichella2015}, building on the work of Pagano and Maalej~\cite{Pagano2013}, designed a taxonomy based on 300 emails and 17 common themes found in app reviews. Their taxonomy consists of four categories—\textit{Information Giving}, \textit{Information Seeking}, \textit{Feature Request}, and \textit{Problem Discovery}—and is particularly relevant for the maintenance and evolution of software systems. These categories help to classify app reviews into categories relevant to software maintenance and evolution.

To the best of our knowledge, this work is the first to identify explanation needs in a real-world company setting and assign these needs to different teams within the company's structure. Additionally, we identify potential existing sources within the company that may already contain the information needed for the explanation or even the explanation itself.

\section{Study Design}
\label{sec:research}

\subsection{Study Concept}

\begin{figure*}[htbp]
    \centering
    \includegraphics[width=0.8\textwidth]{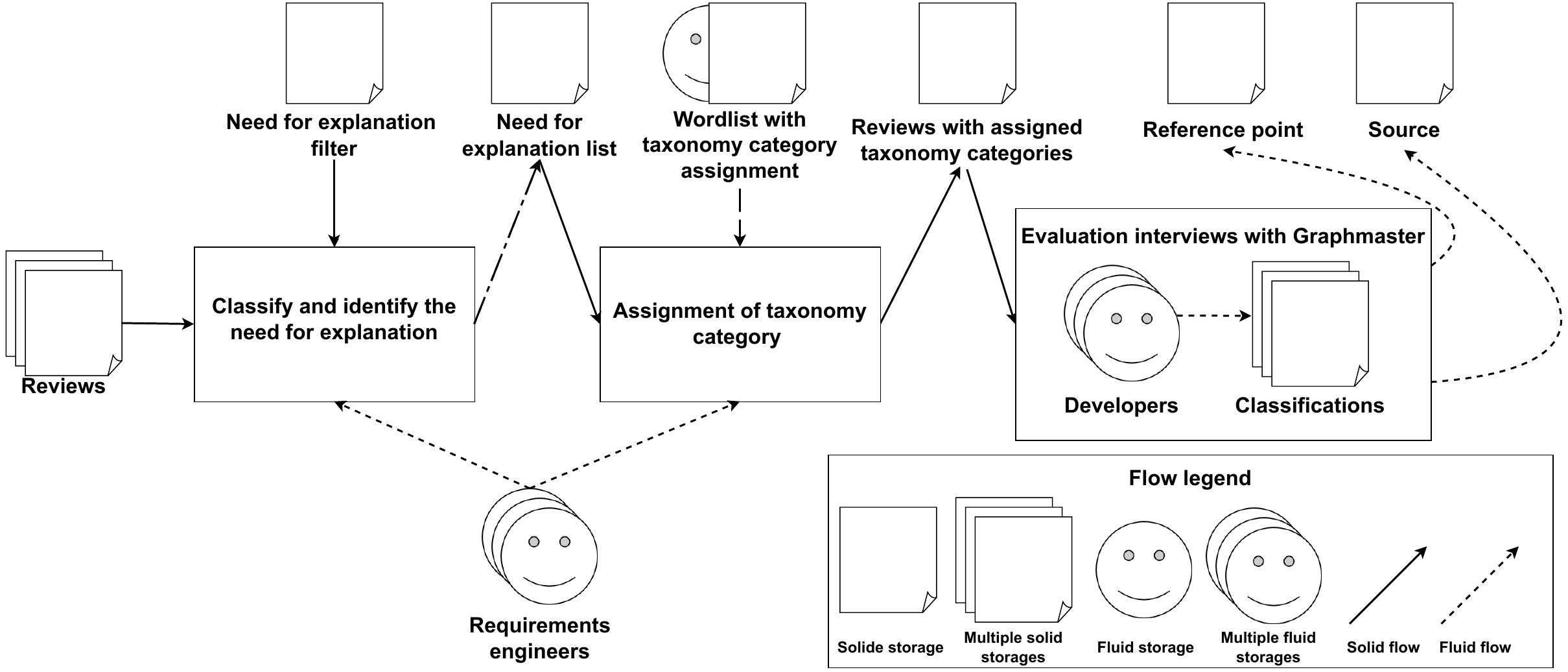}
    \caption{Overview of the study design in the FLOW notation~\cite{stapel2009flow}}
    \label{fig:online-studie}
\end{figure*}

The overall concept of this study, shown in Figure \ref{fig:online-studie}, starts with collecting app reviews from the Google Play Store or Apple App Store. These reviews are then examined regarding the prevalence of explanation needs. If explanation needs are detected using a filter or manual review by a requirements engineer, they are categorized as implicit or explicit. The list of reviews requiring explanations is then assigned to specific taxonomy categories. This process is initially automated using a word and phrase filter and then verified by requirements engineers. Finally, a list of categorized reviews is produced, and through a semi-automated process, reference points and sources for addressing the explanation needs are identified via subsequent interviews and evaluations.

\subsection{Research Goal and Research Questions}

We strive to achieve the following goal, formulated according to the Goal-Definition-Template~\cite{Wohlin.2012}:\\

\setlength{\shadowsize}{2pt}
\noindent 
\shadowbox{ 
\begin{minipage}[t]{0.95\columnwidth} 
\textbf{Research Goal:} \\ \textit{Analyze} the semi-automated management of explanation needs in user reviews of software applications \\
\textit{for the purpose of} improving the efficiency and accuracy of assigning explanation needs \\
\textit{with respect to} enabling scalable and precise categorization and team assignment \\
\textit{from the point of view of} software engineers and customer support teams \\
\textit{in the context of} app reviews. 
\end{minipage} 
}\vspace{0pt}

We address the following research questions:

\begin{itemize}
\item \textbf{RQ1: How accurately can user reviews requiring explanations be categorized into taxonomy-based reference points and sources for providing explanations?} 

Research question \textbf{RQ1} addresses the primary goal of this study. Through interviews and a survey at \textit{Graphmasters GmbH}, we evaluate the assignment of user reviews to a reference point (such as a person or team) and a source (of information). Using taxonomy-based categorization, we assess whether this process can reliably identify appropriate reference points and sources.

\item \textbf{RQ2: Can every user review with explanation needs be assigned a unique reference point and source?} 

\textbf{RQ2} examines whether assigning a unique reference point and source to each review that expresses an explanation need is feasible. It explores potential edge cases and assesses whether ambiguities or overlapping responsibilities arise.

\item \textbf{RQ3: Under what conditions can a company provide a source for a user review, containing information to meet the explainability requirement, and a reference point consisting of an internal team?} 

Research question \textbf{RQ3} focuses on the organizational prerequisites a company must provide to effectively map user reviews requiring explanations to a reference point and source. This includes evaluating internal team structures and the availability of relevant information.

\item \textbf{RQ4: How much of the actual explanation need can a company address?}

The final research question \textbf{RQ4} seeks to assess how much of the actual explanation needs a company can effectively address in practice. This question evaluates the feasibility and scope of explanation management within the constraints of a company's resources.

\end{itemize}

\subsection{Elicitation of Explainability Needs} 

\subsubsection{Graphmasters GmbH}
\textit{Graphmasters GmbH}\footnote{\url{https://www.graphmasters.net/}} is a company specializing in intelligent navigation solutions, focusing on optimizing traffic flow and reducing congestion through real-time route guidance. Based in Hanover, Germany, the company develops apps and systems like NUNAV\footnote{\url{https://www.graphmasters.net/nunav}}, which cater to various use cases such as personal navigation, logistics, and trucking. These applications leverage collaborative routing algorithms to distribute traffic efficiently, minimizing delays and emissions. The company serves diverse industries, including logistics and public transportation, providing innovative solutions to enhance mobility and sustainability. The software can be used on Android and iOS systems, such as cell phones, tablets or Chromebooks.

\subsubsection{Reviews Scraping}

A custom software tool was developed to scrape reviews from the \textit{Google Play Store} and \textit{Apple App Store}. The scraped data included metadata such as star ratings, dates, responses to reviews, review titles, and the text content of the reviews. This data was consolidated into a unified dataset for further analysis.

\subsubsection{Dataset Creation}

The dataset for this study was constructed using user reviews from several apps\footnote{\url{https://www.graphmasters.net/nunav}} developed by Graphmasters GmbH, including the \textit{Nunav Navigation App}, \textit{Nunav Courier App}, \textit{Nunav Trucks App}, and \textit{Nunav Bus App}, sourced from both the Google Play Store and Apple App Store. While the primary focus was on the \textit{Nunav Navigation App}, three additional apps were included to expand the dataset due to their structural similarity. These apps all share core navigation features for road traffic but differ in their primary focus areas, such as logistics. Variations also include minor visual adjustments, settings options, and tailored route optimization features. By including the \textit{Nunav Courier App}, \textit{Nunav Truck App}, and \textit{Nunav Bus App}, the dataset was expanded by 231 reviews, resulting in a total of 2,366 reviews.

\subsubsection{Dataset Analysis}

Explicit and implicit explanation needs in the reviews were identified using a word and phrase filtering method. An approach with large language models was not possible because of the limited training data. The guidelines, description of implicit and explicit explanation needs by Droste et al.~\cite{Droste2024} are available online~\cite{Drostedata2024}. Implicit explanation needs are statements where the user does not pose a direct question but expresses confusion, surprise, or lack of understanding, often using trigger words like "how" or indicating that they cannot use the software without help. In contrast, explicit explanation needs are direct questions or requests where users explicitly express their desire to understand something, interact with the software, or highlight a lack of explanation, often using question words or phrases like "I want to understand.".
The keywords and phrases used to detect explanation needs were collected from various datasets. To improve accuracy, multiple datasets were combined to create a comprehensive filter. An additional classification for "potential explanation need" was introduced during the initial analysis, which was later verified by two requirements engineers.
The largest dataset used for filtering explanation needs was based on Obaidi~\cite{goldstandard-explain-zenodo2024}, which provided 245 distinct phrases. This dataset was expanded by translating the phrases into German using the DeepL translation tool\footnote{https://www.deepl.com/}, as most of the reviews in this study were written in German. The dataset consists of phrases linked to explicit or implicit explanation needs. Additionally, trigger words from Droste et al.~\cite{Drostedata2024} were added to expand the dataset for detecting potential explanation needs. These words were translated into English using DeepL to ensure applicability for reviews in both languages. The dataset was not reduced, even if certain words did not match any reviews, because the dataset had been evaluated in previous studies and might still be applicable to other apps in future analyses.

\begin{table}[!htb]
\center
\setlength{\tabcolsep}{9pt}
\caption{Identified need for explanation according to word and phrase filter of the individual apps}
\label{identifizierter_erklärungsbedarf}

\begin{tabularx}{\columnwidth}{Xrrrr}
\toprule
 & \multicolumn{4}{c}{\textbf{Explanation Need}} \\
 \cmidrule(lr){2-5}
\textbf{App} & \textbf{Explicit} & \textbf{Implicit} & \textbf{Potential} & \textbf{None}\\
\midrule
\textit{nunav navigation} & 24 & 28 & 425 & 1659\\
\textit{nunav truck} & 0 & 0 & 0 & 7\\
\textit{nunav logistics} & 2 & 1 & 49 & 135\\
\textit{nunav bus} & 0 & 1 & 10 & 25\\
\cmidrule(lr){2-5}
{Total} & 26 & 30 & 484 & 1826\\
\bottomrule
\end{tabularx}
\end{table}

The resulting table from the automatic categorization was subsequently analyzed by two requirements engineers independently (authors of this paper) using MaxQDA\footnote{https://www.maxqda.com/}. During this process, reviews that were not detected by the filter list were also considered. The specific segments of the reviews containing explanation needs were manually marked and categorized as either explicit or implicit explanation needs. Any "potential explanation needs" that were automatically flagged by the filter list were re-evaluated and relabeled by the two raters as explicit, implicit, or no explanation need.
In cases where discrepancies arose between the two engineers’ categorizations, the distinctions between implicit and explicit explanation needs were carefully evaluated. To resolve these differences, the two requirements engineers discussed the discrepancies to reach a consensus. The final results of this manual labeling process are considered the "ground truth".
The manual labeling identified 158 explanation needs across the 2,366 reviews. The discrepancy between the number of needs (158) and the reviews containing them (148) is due to 10 reviews containing more than one explanation need. Specifically, five reviews initially labeled by the filter list as having explicit or potential explanation needs were found to contain multiple needs, leading to their multiple counts.

\subsection{Evaluation of Explainability Need} 

To evaluate the data, interviews and surveys were conducted with Graphmasters. The participants were asked to assign app reviews to taxonomy categories and identify the teams within the company responsible for addressing the explanation needs. Four members of Graphmasters' support team were recruited for the study. As the team responsible for responding to user reviews, they represented a group of experts well-suited for this task. The in-person interviews involved assigning reviews with explanation needs to both taxonomy categories and reference points within the organization.

An example is as follows:\noindent

\setlength{\shadowsize}{2pt}
\noindent
\shadowbox{
\begin{minipage}[t]{0.95\columnwidth}
\textbf{Question}: \textit{Why does the app sometimes vibrate?}

\noindent
\textbf{Classification group}: \textit{Mobile}

\noindent
\textbf{Classification taxonomy category}: \textit{Unexpected system behavior}
\end{minipage}
}\vspace{0pt}

After the interviews, participants completed an online survey, in which additional app reviews were categorized using the same structure of questions and answer options as in the interviews. The interviews served as an introduction to the subject and clarified any initial misunderstandings. To ensure consistency across all interviews, detailed interview guidelines were prepared. The online survey also allowed participants to provide additional comments if they wished to clarify their decisions. During the interviews, 75 reviews were analyzed in total. They were divided into three sets of 25 reviews each. The four participants were split into two groups: both groups evaluated the first set of 25 reviews, while the second set was assigned to the first group and the third set to the second group. This structure allowed the interviews to remain within a 30-minute time frame while covering all 75 reviews.
The remaining 83 reviews, out of the 158 reviews identified as having explanation needs, were evaluated through the online survey.

\subsection{List of Reference Points and Source}

\subsubsection{Reference Point}
\label{bezugspunkt}
To automatically assign a reference point for addressing the explanation needs identified in the reviews, taxonomy categories are employed. Each taxonomy category is mapped to one or more reference points within the company. The data for this assignment was gathered from the interviews with the Graphmasters support team. If the mapping of a team to a taxonomy category is not clear, a hierarchical structure is created, ranking the teams based on their likelihood of providing a suitable answer to the explanation need. Any team associated with at least 25\% of a taxonomy category is included in this hierarchy. For example, if a team has been assigned to a taxonomy category at least 1/4 of the total assignments to this taxonomy category, the team is included in the hierarchy of possible team assignments for the taxonomy category.

\subsubsection{Source}

The source containing the answer to the explanation need is determined using an API from the Graphmasters support website. The script, which uses the API, compares the similarity between support articles on the Graphmasters support page and the user’s query using the SequenceMatcher from Python’s difflib module\footnote{\url{https://docs.python.org/3/library/difflib.html}}. The support articles on the Graphmasters support page were created manually by Graphmasters support members for general explanatory purposes. This comparison helps identify whether a specific support article addresses the question at hand. If no relevant support article is found, the script retrieves all answers from responses provided to the review in the app store. These responses must then be manually read and assigned. If no suitable answer is found there either, a new response to the explanation need is formulated by Graphmasters.

\subsection{Assignment of Reference Points and Source} 

The reference point was determined based on the evaluation of interviews and surveys conducted with Graphmasters, in which the explanation needs were mapped to internal teams within the company. Additionally, the taxonomy categories assigned by the requirements engineers were utilized. If a team received more than 25\% of the total assignments for a particular taxonomy category, that team was automatically assigned to the category by the script (see Table \ref{taxonomiekategorie_gruppenzuordnung}).

The source for responding to explanation needs was determined through three different sources: the Graphmasters support website via an API, responses to Google Play Store reviews provided by Graphmasters, and manual responses when neither source provided a suitable answer. The hierarchy prioritized the support website, followed by the Google Play Store responses, and finally manual drafting of new responses.

The API identified 27 support articles that could address explanation needs. Of these, 10 articles were deemed irrelevant to the explanation needs. Among the 17 valid support articles, 11 had corresponding review responses in the Google Play Store. As of August 11, 2024, Graphmasters has published a total of 69 combined support articles across the evaluated apps. However, some articles are duplicated, such as those addressing the same question across multiple apps. For example, "Which languages does the NUNAV (Courier/Cargobike/Navigation/Trucks) app support?" All these questions point to the same support article. Additionally, Graphmasters publishes release notes for each version of the apps, which can be used as responses to explanation needs, especially when features are added, bugs are fixed, or new explanations are introduced in the system.

In addition, 126 further responses to explanation needs were sourced from the Google Play Store. The dataset was manually expanded with 15 new responses drafted by Graphmasters to address outstanding explanation needs. Of these, 13 responses were for Apple App Store reviews, which accounted for only 18 of the total 158 explanation needs.

\subsection{Addressability of Explanation Needs} 

To answer RQ4, the aim was to investigate how much of the identified explanation needs could be addressed by Graphmasters. To achieve this, the list of identified explanation needs was provided to a support team member, who indicated which explanation needs had already been solved, would be solved, or could not be solved due to various reasons.

\subsection{Performance Metrics}
In order to measure the accuracy of the labeling in comparison to the ground truth, we considered \textit{Precision}, \textit{Recall}, \textit{F1-Score}, and \textit{Accuracy}.

\subsection{Interrater Agreement}
The classification of the taxonomy category was carried out by two requirements engineers, and the interrater agreement was calculated using Cohen's kappa. After a final agreement was reached between the two requirements engineers, this was regarded as ground truth for the remainder of the study.

The interviews and surveys provided insights into the practical applicability of the approach in a real-world setting. The interrater reliability was measured using the Kappa statistic, specifically Fleiss' $\kappa$~\cite{fleiss1971measuring} and Cohen's $\kappa$~\cite{cohen1960agreement}, which indicate the degree of agreement between participants. Since the reviews with explanation needs were evaluated by varying numbers of participants, individual Kappa values were calculated for review sets 1-25, 26-50 and 51-75, as well as the validity of the correct taxonomy category assignment (see \ref{taxonomiekategorie_umfrage}, \ref{taxonomieoberkategorie_umfrage}). The four participants were divided into two groups for the interviews (group 1 and group 2). For reviews 1-25 both group 1 and group 2 were interviewed and evaluated with Fleiss' Kappa. Cohen's Kappa was calculated for reviews 26-50 and 51-75, because only 2 participants were interviewed for each review range. No Kappa value was assigned to reviews 76-158, as these were evaluated by a single participant (Group 3), making interrater reliability irrelevant. Additionally, another table was used to evaluate the categorization into overarching taxonomy categories, which were derived from the individual taxonomy categories after the fact (see \ref{taxonomieoberkategorie_umfrage}).

\section{Results}
\label{sec:results}

\subsection{Detection of Explanation Needs}

The word and phrase filtering system labeled 548 review--need instances as containing some form of explanation need. The filter was designed to capture as many explanation needs as possible. The results are in Table \ref{evaluation_erklärungsbedarf}, the overall results in Table \ref{metrik_gelabelter_erklärungsbedarf}. When summing all types of labels (explicit, implicit, and potential explanation needs), the system achieved a recall of 0.8291 (\ref{metrik_gelabelter_erklärungsbedarf}). However, the precision of the combined labels was low at 0.2391, resulting in a modest F1-score of 0.3533. This low precision was accepted to prioritize achieving a high recall. The pre-filtering process reduced the dataset from 2,376 review--need instances to 548, with an overall accuracy of 0.3149. The low accuracy is primarily attributed to the precision of 0.2106 for detecting potential explanation needs.
The precision for detecting explicit explanation needs was significantly higher, reaching 0.8276. This can be attributed to the filtering system being specifically designed to capture explicit needs with high precision. However, this came at the cost of a lower recall of 0.1765. The detection of implicit explanation needs showed the lowest values, as implicit needs often do not present directly as questions. The precision for implicit needs was 0.1333, with a recall of 0.1818, resulting in an F1-score of 0.1538.

\begin{table}[!htb]
\centering
\setlength{\tabcolsep}{9pt}
\caption{Explanation need classifications from a filter list in relation to the ground truth (explanation needs labeled by requirements engineers).}
\label{evaluation_erklärungsbedarf}
\begin{tabularx}{\columnwidth}{Xrrrr}
\toprule
 & \multicolumn{4}{c}{\textbf{Explanation Need}} \\
 \cmidrule(lr){2-5}
\textbf{Metric} & \textbf{Explicit} & \textbf{Implicit} & \textbf{Potential} & \textbf{None}\\
\midrule
Precision & 0.8276 & 0.1333 & 0.2106 & 0.9858 \\
Recall & 0.1765 & 0.1818 & 0.6519 & 0.8124\\
F1-Score & 0.2909 & 0.1538 & 0.3184 & 0.8901\\
\bottomrule
\end{tabularx}
\end{table}

\begin{table}[!htb]
\centering
\setlength{\tabcolsep}{9pt}
\caption{Overall classified explanation need from a filter list compared to ground truth (explanation needs labeled by requirements engineers).}
\label{metrik_gelabelter_erklärungsbedarf}
\begin{tabularx}{\columnwidth}{XXXX}
\toprule
\textbf{Precision} & \textbf{Recall} & \textbf{F1-Score} & \textbf{Accuracy} \\
\midrule
0.2391 & 0.8291 & 0.3533 & 0.2282 \\
\bottomrule
\end{tabularx}
\end{table}

\subsection{Detecting Taxonomy Categories Using a Word Filter}

Using a fine-grained word filter to detect taxonomy categories, after adjustments, resulted in a precision of 0.6727 and a recall of 0.2341. In comparison, a broad word filter, which included more words but lacked precision in assigning words to taxonomy categories, achieved a precision of 0.5079 and a recall of 0.4051. When comparing F1-scores, the broad word filter outperformed the fine filter, with an F1-score of 0.4507 compared to 0.3473 for the fine filter. Since precision is more relevant for identifying the correct reference point and reducing the need for corrections, the F-score was adjusted using a $\beta$-value of 0.2 to prioritize precision. With this adjustment, the fine filter achieved an F-score of 0.5126, while the broad filter reached 0.4873.

Assigning taxonomy categories based solely on individual words has limited reliability. When requirements engineers attempted to categorize reviews using only keywords, the precision was 0.2211, and the recall 0.2857, indicating poor performance. Due to these suboptimal results, this method is not practical in real-world applications.

Both the fine and broad word filters were developed based on an evaluation of reviews that had already been categorized with explanation needs. These filters are not suitable for use with new apps and can only be applied after the taxonomy categories for other reviews of an already evaluated app have been established.

\begin{table}[!htb]
\centering
\setlength{\tabcolsep}{9pt}
\caption{Evaluation of word filters for assignment to taxonomy categories.}
\label{evaluation_wörterfilter_erklärungsbedarf}

\begin{tabularx}{\columnwidth}{Xrrr}
\toprule
\textbf{Metrics} & \textbf{Classification REs} & \textbf{Coarse filter} & \textbf{Fine filter} \\
\midrule
Precision & 0.2211 & 0.5079 & 0.6727 \\
Recall & 0.2857 & 0.4051 & 0.2341 \\
F1-Score & 0.2490 & 0.4507 & 0.3473 \\
Accuracy & 0.2857 & 0.4051 & 0.2341 \\
\bottomrule
\end{tabularx}
\end{table}

\subsection{Taxonomy Category Assignment}
The interrater agreement between the two requirements engineers when classifying the taxonomy categories resulted in a kappa value of 0.61, which we interpret as substantial agreement according to Landis and Koch~\cite{landis1977measurement}. Most disagreements were regarding the \textit{Business} category. 21 out of 55 were between \textit{Business} and \textit{Meta information}. Many reviews contained questions about the platform. For example, one review stated “I find the upcoming motorcycle version exciting - what will make it different?”. One requirements engineer was unsure and labeled \textit{Meta information}, the other requirements engineer labeled \textit{Business}, because it is a question which can be answered by the company. In almost all of these cases, the agreed label was \textit{Business}.
The second highest number of disagreements (12) are in the \textit{Tutorial} category. This is due to the fact that users in their reviews did not ask for a tutorial, but one requirements engineer often interpreted that a tutorial was desired. For example, the review “I would have liked to know whether it can also be used offline.” was labeled \textit{Tutorial} by one requriement engineer.

The analysis of Kappa values for taxonomy category assignment by Graphmasters developers revealed low to moderate agreement. Significant differences were noted in the level of agreement between the groups. For reviews 1-25, where both Group 1 and Group 2 were involved, a Kappa value of 0.21 was calculated, indicating marginal agreement. Group 1's evaluation of reviews 26-50 yielded a Kappa value of 0.39, which falls into the category of fair agreement but approaches the 0.40-0.60 range defined by Landis and Koch~\cite{Landis1977}, which is considered moderate agreement. In contrast, the agreement for reviews 51-75 was significantly lower, with a Kappa value of 0.004, indicating almost no agreement.

When the taxonomy categories were aggregated into broader categories, the Kappa values improved across all sets of reviews. The most significant increase was observed for reviews 1--25, where Fleiss' $\kappa$ rose from 0.202 to 0.442, indicating moderate agreement. Aggregating taxonomy categories improved agreement for reviews 1-25 and 26-50 from marginal to moderate levels, while the agreement for reviews 51-75 remained poor. In terms of validity, the participants' categorization accuracy for taxonomy categories was less than 50\%. However, when broader taxonomy categories were considered, the validity improved to between 54\% and 72\%.

\begin{table}[!htb]
\centering
\setlength{\tabcolsep}{3pt}
\caption{Taxonomy assignment of interviews and survey to subcategories.}
\label{taxonomiekategorie_umfrage}

\begin{tabularx}{\columnwidth}{XXXrrr}
\toprule
\textbf{IDs} & \textbf{$\boldsymbol{n}$} & \textbf{Group} & \textbf{Validity} & \textbf{Cohen's $\kappa$} & \textbf{Fleiss' $\kappa$} \\
\midrule
1--25   & 4        & 1,2    & 37\%   & /     & 0.202 \\
26--50  & 2        & 1      & 42\%   & 0.393 & / \\
51--75  & 2        & 2      & 38\%   & 0.004 & / \\
76--158 & 1        & 3      & 26.51\%& /     & / \\
\bottomrule
\end{tabularx}
\end{table}

\begin{table}[!htb]
\centering
\setlength{\tabcolsep}{3pt}
\caption{Taxonomy assignment of interviews and survey to superordinate categories.}
\label{taxonomieoberkategorie_umfrage}

\begin{tabularx}{\columnwidth}{XXXrrr}
\toprule
\textbf{IDs} & \textbf{$\boldsymbol{n}$} & \textbf{Group} & \textbf{Validity} & \textbf{Cohen's $\kappa$} & \textbf{Fleiss' $\kappa$} \\
\midrule
1--25   & 4        & 1,2    & 62\%   & /     & 0.442 \\
26--50  & 2        & 1      & 72\%   & 0.518 & / \\
51--75  & 2        & 2      & 54\%   & 0.085 & / \\
76--158 & 1        & 3      & 43.37\%& /     & / \\
\bottomrule
\end{tabularx}
\end{table}

\subsection{Team Assignments}
The team assignment agreement for the reviews (\ref{taxonomiekategorie_reviews}) followed a similar pattern to the taxonomy category assignments. The highest Kappa value of 0.558 was observed for reviews 26-50, indicating moderate agreement. Reviews 1-25 showed a Kappa value of 0.307, which is considered fair agreement, while reviews 51-75 again demonstrated poor agreement with a Kappa value of 0.146. As in previous sections, reviews 76-158 did not receive a Kappa value, as they were evaluated by only one participant.

\begin{table}[!htb]
\centering
\setlength{\tabcolsep}{11pt}
\caption{Taxonomy assignment of reviews}
\label{taxonomiekategorie_reviews}
\begin{tabularx}{\columnwidth}{XXXrrr}
\toprule
\textbf{IDs} & \textbf{$\boldsymbol{n}$} & \textbf{Group} & \textbf{Cohen's $\kappa$} & \textbf{Fleiss' $\kappa$} \\
\midrule
1--25   & 4 & 1,2 & / & 0.307 \\
26--50  & 2 & 1   & 0.558 & / \\
51--75  & 2 & 2   & 0.146 & / \\
76--158 & 1 & 3   & / & / \\
\bottomrule
\end{tabularx}
\end{table}

\subsection{Reference Points and Source}

By applying the 25\% threshold for team inclusion in the taxonomy category assignment, an average validity of 79.2\% was achieved. This calculation considers cases where a team was ranked as the second or third option but was still considered a correct match. Given the low to moderate agreement in interrater reliability, which indicates some ambiguity in the assignment process, a hierarchical approach was adopted instead of direct assignment to a single team.  The teams \textit{Mobile} and \textit{Support} were frequently assigned, with \textit{Mobile} being the top-ranked team eight times and \textit{Support} six times. On average, the first-ranked team provided the correct answer to the explanation need 52.5\% of the time, the second option 34.9\% of the time, and the third option (in the one case where it was used) 28.3\% of the time.

The category \textit{Meta} was assigned 25 times out of the 283 evaluations. This "team" serves as a fallback option when no clear team assignment could be made.

\begin{table}[!htb]
\centering
\setlength{\tabcolsep}{9pt}
\caption{Taxonomy~\cite{Droste2024} category team assignment.}
\label{taxonomiekategorie_gruppenzuordnung}

\begin{tabularx}{\columnwidth}{Xlr}
\toprule
\textbf{Taxonomy category} & \textbf{Team(s)} & \textbf{Applicability} \\
\midrule
\multirow{3}{*}{Business} 
    & 1. Business & 28\% \\
    & 2. Support & 28\% \\
    & 3. Mobile & 28\% \\
\hline
\multirow{2}{*}{Operation}
    & 1. Mobile & 43\% \\
    & 2. Support & 41\% \\
\hline
\multirow{2}{*}{Tutorial}
    & 1. Support & 54\% \\
    & 2. Mobile & 30\% \\
\hline
\multirow{2}{*}{Navigation}
    & 1. Support & 35\% \\
    & 2. Mobile & 35\% \\
\hline
\multirow{2}{*}{Algorithms}
    & 1. Routing & 42\% \\
    & 2. Support & 28\% \\
\hline
Consequences
    & Mobile & 88\% \\
\hline
\multirow{2}{*}{Unexpected system behavior}
    & 1. Mobile & 42\% \\
    & 2. Routing & 33\% \\
\hline
Bugs \& Crashes
    & Mobile & 57\% \\
\hline
\multirow{2}{*}{User Interface}
    & 1. UI/UX & 55\% \\
    & 2. Mobile & 36\% \\
\hline
\multirow{2}{*}{Privacy}
    & 1. Mobile & 75\% \\
    & 2. Meta & 25\% \\
\hline
Security
    & n.a. & n.a. \\
\hline
\multirow{2}{*}{Meta information}
    & 1. Mobile & 36\% \\
    & 2. Support & 36\% \\
\hline
Terminology
    & n.a.  & n.a.  \\
\hline
\multirow{2}{*}{System-specific elements}
    & 1. Support & 75\% \\
    & 2. Mobile & 25\% \\
\bottomrule
\end{tabularx}
\end{table}

\subsection{Addressable Results}

Of the 158 identified explanation needs, 139 (88\%) were successfully resolved. The remaining needs could not be fully resolved by the company. Common reasons for unresolved explanation needs include cases that require further inquiries, or issues unrelated to any of Graphmasters' apps. However, Graphmasters intends to provide a response to every user with an explanation need.

\section{Discussion}
\label{sec:discussion}

In the following, we answer the research questions, present threats to validity, and interpret the results.

\subsection{Answering the Research Questions}
\label{sec:beantworten-der-forschungsfragen}

\textbf{RQ1: How accurately can user reviews requiring explanations be categorized into taxonomy-based reference points and sources for providing explanations?} 

By adopting a hierarchical approach to reference points—where multiple potential teams are ranked by their likelihood of providing a relevant answer—the correct team could be assigned in 79.2\% of cases. Without this hierarchy, assigning a single team resulted in a success rate of 52.5\%. The teams \textit{Mobile} and \textit{Support} were the most frequently assigned, with \textit{Mobile} being the top team in eight instances and \textit{Support} in six. Unlike reference points, sources are not directly linked to taxonomy categories, as the taxonomy only indicates an area of need. However, the taxonomy helps identify the origin of the need, from which the appropriate source can be derived. Sources are identified through the support API and existing review responses in the Google Play Store.

\textbf{RQ2: Can every user review with explanation needs be assigned a unique reference point and source?} 

\noindent In the interviews and survey, 258 out of 283 reference points were successfully assigned to an explanation need. As with taxonomy categories, there were exceptions, which were labeled with the tag \textit{Meta}.
At Graphmasters, however, such ambiguities do not arise in practice, as the support team ultimately handles all \textit{Meta} cases.

Graphmasters provided responses to all explanation needs identified. However, not all sources were initially available. For 15 of the 158 explanation needs, Graphmasters formulated new responses, as there were neither relevant articles on the Graphmasters support website nor existing responses from the Google Play Store. For reviews from the Apple App Store, the only available source was the support website, which significantly limited the options for finding relevant sources. Of the 15 newly formulated responses, 13 were created for Apple App Store reviews, which accounted for 18 of the total 158 explanation needs.

\textbf{RQ3: Under what conditions can a company provide a source for a user review, containing information to meet the explainability requirement, and a reference point consisting of an internal team?} 

For a company to assign a reference point to an explanation need, it must have a clear internal structure that outlines which teams are responsible for specific areas. Additionally, a dataset linking teams to explanation needs must already exist, allowing for future needs to be inferred from past assignments. This dataset can be developed through interviews, surveys, or other evaluation methods. Once these foundations are established, a reference point can be automatically assigned to an explanation need.

To provide a source for responding to explanation needs in user reviews, companies can create general articles on their support website, offering pre-formulated responses to recurring needs. Additionally, companies can publicly respond to user reviews in app stores, either manually or through automated processes. If no source is available, the company can provide a new response. In cases where the explanation need is not directly related to the company or lacks sufficient user-provided information, the company may be unable to offer a source. For non-company-related questions, Graphmasters typically provides referral responses rather than detailed answers due to the limited characters in the review answers.

\textbf{RQ4: How much of the actual explanation need can a company address?}

Graphmasters was unable to fully resolve 19 (12\%) out of 158 explanation needs because they either fell outside the company's scope or lacked sufficient information to resolve the issue. For example, a vague request like, "Sometimes weird lines appear on the map," or a question unrelated to the company's apps, such as, "How do I get an update for my VW Passat navigation system?" could not be resolved. However, Graphmasters intends to provide a response to every user with an explanation need.

\subsection{Interpretation}
\label{sec:interpretation}

\subsubsection{Taxonomy Customization}
\label{taxonomieerweiterung}
During the assignment of taxonomy categories by the two requirements engineers, it became apparent that not every explanation need could be categorized using the taxonomy proposed by Droste et al.~\cite{Droste2024}, as was also discussed in their original publication.

To allow for more specific categorization, the taxonomy was expanded to include the categories \textit{Business} and \textit{Meta Information}, similar to the approach taken by Obaidi~\cite{goldstandard-explain-zenodo2024}. Although \textit{Meta Information} was initially included in Droste et al.'s dataset~\cite{Drostedata2024}, it was later removed as many needs were reclassified into other categories. In this study, both \textit{Business} and \textit{Meta Information} were treated as standalone categories. The \textit{Business} category captures explanation needs that are not directly related to the system itself but pertain to questions about the provider. The \textit{Meta Information} category was used to collect explanation needs that could not be clearly assigned to other categories or spanned multiple categories. Ultimately, the \textit{Meta Information} category accounted for 18 of the 158 explanation needs.

Additionally, 10 explanation needs were classified under a new category, \textit{Feature Questions}. This category, which was not derived from previous studies, covers cases where multiple taxonomy categories are impacted, making clear assignment difficult. For example, a review asking, "Does the app support Polish and other languages?" touches on the \textit{Operation} category for functionality, the \textit{Introduction} category for onboarding, and the \textit{Business} category regarding future plans. The \textit{Feature Questions} category allows for clearer categorization in such cases.

In addition, the \textit{Tutorial} category was not clear, as according to the guidelines the desire for a tutorial describes this category. However, the labeling was often interpreted as meaning that the user wanted a tutorial, although it was not mentioned. Since the idea of this category often differs from the actual description in the guidelines, we suggest deleting this category. 

\subsubsection{Categorization into a Taxonomy}
The validity of the taxonomy category assignments (see \ref{taxonomiekategorie_umfrage}), based on the participants' evaluations, indicates challenges in assigning categories. These challenges arose either from a lack of complete understanding or from subjective decision-making based on practical experience, which sometimes misaligned with the core explanation need. During the interviews, it became evident that the number of taxonomy categories was overwhelming, making it difficult for participants to keep track. Reducing the number of categories to only broader categories and standalone categories could improve the validity of the assignments by 23.6\% on average (see \ref{taxonomieoberkategorie_umfrage}). However, this simplification could negatively affect the precision of assigning reference points to explanation needs, as these assignments depend on the specific taxonomy categories.

\subsubsection{Assignment to a Reference Point}
The assignment of reference points among the participants was inconsistent, with Kappa values ranging from 0.146 to 0.558, indicating that clear team assignments could not be made for many explanation needs. To address this, a probability-based ranking system for reference point assignment was developed. The variance in responses can likely be attributed to subjective interpretations of how explanation needs should be handled. In practice, many participants selected the support team as the reference point, as it typically handles explanation needs. However, other participants chose different teams that could potentially address the explanation need outside the support team. This discrepancy between theory and practice, as revealed in the interviews, played a significant role in the inconsistent assignments. Despite clear instructions to name the original team responsible for addressing the explanation need, participants often defaulted to their practical experience, which skewed the categorization. Therefore, in practice, it is challenging to accurately assign a team that theoretically knows the answer to the explanation need.

\subsubsection{Relevance of Addressing Explanation Needs in Industry}

The findings of this study highlight the practical relevance of explanation needs in industry settings. Our results show that Graphmasters was able to address or respond to 88\% of the identified explanation needs, underscoring the importance of systematically identifying and responding to these needs. This high percentage demonstrates that explanation needs are not merely theoretical constructs but play a significant role in improving user satisfaction and ensuring product success. Addressing these needs directly impacts how users interact with the app and perceive its usability, contributing to a better overall user experience.
Moreover, the remaining 12\% of the explanation needs, which the company could not fully resolve, also offer valuable insights. Even in these cases, the company committed to providing users with a response, despite not being able to offer a complete explanation. This reflects a broader industry practice of maintaining open communication with users and managing their expectations, even when a full resolution is not possible. By ensuring that every user inquiry is met with some form of response, companies like Graphmasters demonstrate their commitment to customer service and transparency, which can foster trust and long-term user loyalty.
These findings highlight that addressing explanation needs is not only about solving technical issues but also about ensuring that users feel heard and supported. The process of identifying, categorizing, and addressing these needs is vital for improving the quality of software and meeting user expectations in an increasingly competitive marketplace.

This study contributes to the growing body of research on explainability in software engineering by providing a practical framework that combines automated categorization with manual oversight. It also highlights the importance of explainability as a non-functional requirement, especially in user-facing software products. The expanded taxonomy serves as a guide for other researchers and practitioners aiming to categorize user feedback based on explainability needs.

\subsection{Future Work}
\label{sec:ausblick}

Building on the findings of this study, several areas for future research and improvement have been identified:

The current approach assigns different internal teams as a reference point for addressing explanation needs. However, future work could refine this process by identifying individual experts based on code ownership or task division, thus providing more targeted assignments for explanation requests.
Additionally, larger datasets could be used to train artificial intelligence models capable of automatically assigning both a reference point and a relevant source for addressing explanation needs. This would enhance the automation and accuracy of the system.
Further evaluation of the taxonomy used for categorizing explanation needs could help refine or even eliminate the current \textit{Meta Information} category. Exploring the applicability of alternative taxonomies could also provide valuable insights into improving classification accuracy.
In this study, team assignments were based on input from the support team at Graphmasters, who mapped the explanation needs to internal teams. However, the accuracy of these team assignments in resolving the explanation needs has not been evaluated. Future research could investigate whether the assigned teams are indeed the most suitable for resolving the issues.
Finally, expanding the dataset to include a larger company with a more diverse internal structure would provide an opportunity to assess the generalizability of this approach and explore its scalability in more complex organizational settings.

\subsection{Threats to Validity}
\label{sec:validiteat}
Wohlin et al.~\cite{Wohlin2012} identified four categories of threats that may affect the validity of methods and data used in research studies. These categories are applied and analyzed in the context of this study.

\subsubsection{Construct Validity}
When attempting to classify reviews using an unfamiliar taxonomy, the complexity of the taxonomy can lead to misinterpretations and improper usage. Misunderstanding the taxonomy can result in incorrect categorizations, which may affect the outcomes of the study. To mitigate this threat, participants were given sufficient reading time and were allowed to ask clarification questions at the beginning of the interviews. The taxonomy categories were made available throughout the interview, and examples for each category were provided to ensure clarity and accurate application. The word and phrase filtering system for detecting explanation needs is based on predefined datasets and rules, which might not capture the full range of expressions used by users. This could lead to biased or incomplete categorization of explanation needs. To mitigate this threat, we also used a filter with high recall, but low precision, to find all explanation needs.

\subsubsection{Internal Validity}
The timing and context of the interviews could introduce biases that affect the participants' responses. For instance, if interviews are held during participants' personal time, they may rush through the process, negatively impacting data quality. To address this threat, all interviews were conducted during working hours at the participants' workplace. This ensured that the time spent on the interview was consistent and fair across all participants. 
The app reviews analyzed in this study are likely to evolve as the apps are updated and as new features are introduced. The study does not account for temporal changes in app functionality, which may affect the relevance of reviews over time. We consider this threat acceptable, because the study focused on a static snapshot of reviews. While app reviews may evolve, the automated process itself remains flexible and can be reapplied periodically to capture future changes in user feedback.

\subsubsection{Conclusion Validity}
The limited sample size of four participants poses a potential threat to conclusion validity. This small sample may limit the statistical significance of the data, leading to potentially inaccurate conclusions. To address this, participants were selected based on their expertise. All interviewees had at least three years of experience working at the company, ensuring that the interviews captured relevant and knowledgeable insights. Additionally, an interview guideline was developed to maintain consistency across interviews. This guideline structured the interview process, specifying when explanations were provided and when participants were expected to derive solutions independently.

\subsubsection{External Validity}
Ensuring external validity was challenging due to the limited sample size and the focus on a single company. The results of this study may not generalize to other organizations or software systems, as the study was conducted within the specific context of Graphmasters GmbH, which operates a specialized navigation app. To address this threat, consistent conditions were maintained for all participants. Each participant had at least three years of experience within the company and was either part of the support team or possessed the skills required for such a role. However, given the narrow scope of the company’s focus, the findings of this study may not be directly applicable to other contexts or industries.
Our study did not include a longitudinal component, which presents a threat to external validity. The proposed process has not been tested over time, meaning that we cannot be sure how well it adapts to changes in app functionality or user feedback.

\section{Conclusion}
\label{sec:conclusion}
This paper investigated the semi-automated management of explanation needs in user reviews, focusing on classifying reviews into taxonomy categories and assigning appropriate reference points and sources. A combination of manual and automated processes was employed to streamline the task, leveraging a dataset of app reviews from Graphmasters GmbH.
The results showed that by employing a hierarchical approach, the correct team could be assigned to address explanation needs in 79.2\% of the cases. Without this hierarchy, the success rate for assigning a single team dropped to 52.5\%. A total of 139 out of 158 explanation needs were addressed or received a response, with 15 requiring newly formulated responses due to the lack of pre-existing sources. Notably, 13 of these new responses were for Apple App Store reviews, highlighting limitations in the availability of support articles across platforms.
The taxonomy used in this study was expanded to include new categories such as \textit{Business} and \textit{Meta Information}, allowing for more specific categorization of explanation needs that were previously uncategorized. 

However, challenges remained in detecting taxonomy categories using word filters, with broad filters achieving better recall but lower precision compared to fine filters. Interrater agreement on taxonomy category assignment also posed challenges, though it improved when broader categories were used.
Despite these difficulties, the semi-automated process reduced the manual workload and proved effective in addressing user explanation needs with sufficient accuracy, recall, and F1-score. Future work should focus on refining the taxonomy for greater accuracy, exploring the use of artificial intelligence for further automation, and enhancing the system’s ability to identify both reference points and sources with minimal manual intervention.

\section*{Acknowledgment}
This work was funded by the Deutsche Forschungsgemeinschaft (DFG, German Research Foundation) under Grant No. 470146331, project softXplain (2022-2025).
We thank Graphmasters GmbH (Hannover, Germany) for allowing us to conduct this study in their company.

\bibliographystyle{IEEEtran}
\bibliography{references.bib}

@inproceedings{unterbusch2023explanation,
  title={Explanation Needs in App Reviews: Taxonomy and Automated Detection},
  author={Unterbusch, Max and Sadeghi, Mersedeh and Fischbach, Jannik and Obaidi, Martin and Vogelsang, Andreas},
  booktitle={2023 IEEE 31st International Requirements Engineering Conference Workshops (REW)},
  pages={102--111},
  year={2023},
  organization={IEEE}
}

@article{landis1977measurement,
  title={The measurement of observer agreement for categorical data.},
  author={J. Landis and G. Koch},
  journal={Biometrics},
  volume={33 1},
  pages={159-74},
  year={1977},
  doi={10.2307/2529310},
  publisher = {International Biometric Society, Wiley}
}

@article{cohen1960agreement,
  author={Jacob Cohen},
  title={A Coefficient of Agreement for Nominal Scales},
  journal={Educational and Psychological Measurement},
  volume={20},
  number={1},
  pages={37-46},
  year={1960},
  doi={10.1177/001316446002000104}
}

@article{fleiss1971measuring,
  title={Measuring nominal scale agreement among many raters.},
  author={Fleiss, Joseph L},
  journal={Psychological Bulletin},
  volume={76},
  number={5},
  pages={378-382},
  year={1971},
  doi={10.1037/h0031619},
  publisher={American Psychological Association}
}

@book{Wohlin.2012,
 year = {2012},
 title = {Experimentation in software engineering},
 address = {Berlin},
 publisher = {Springer},
 author = {Claes Wohlin and Runeson, Per and Höst, Martin and Ohlsson, Magnus C. and Regnell, Björn and Wesslén, Anders},
 isbn = {9783642290442},
 numpages = {236},
 doi = {10.1007/978-3-642-29044-2},
}

@INPROCEEDINGS{stapel2009flow,
  author={Stapel, Kai and Knauss, Eric and Schneider, Kurt},
  booktitle={2009 Collaboration and Intercultural Issues on Requirements: Communication, Understanding and Softskills}, 
  title={Using FLOW to Improve Communication of Requirements in Globally Distributed Software Projects}, 
  year={2009},
  volume={},
  number={},
  pages={5-14},
  doi={10.1109/CIRCUS.2009.6}
}

@misc{goldstandard-explain-zenodo2024,
  author       = {Obaidi, Martin},
  title        = {{Dataset: Gold standard dataset for explainability 
                   need detection in app reviews.}},
  month        = sep,
  year         = 2024,
  publisher    = {Zenodo},
  doi          = {10.5281/zenodo.11522828},
  url          = {https://doi.org/10.5281/zenodo.11522828}
}

@article{Chazette2021,
  author = {L. Chazette and W. Brunotto and T. Speith},
  title = {Exploring explainability: a definition, a model, and a knowledge catalogue},
  journal = {29th international requirements engineering conference},
  year = {2021},
  volume = {29},
  pages = {197-208},
  doi = {10.1109/RE51729.2021.00025},
  language = {English},
  publisher = {IEEE}
}

@INPROCEEDINGS{Droste2024,
  author={Droste, Jakob and Deters, Hannah and Obaidi, Martin and Schneider, Kurt},
  booktitle={2024 IEEE 32nd International Requirements Engineering Conference (RE)}, 
  title={Explanations in Everyday Software Systems: Towards a Taxonomy for Explainability Needs}, 
  year={2024},
  volume={},
  number={},
  pages={55-66},
  doi={10.1109/RE59067.2024.00016}
}

@article{Chazette2020,
  author = {L. Chazette and K. Schneider},
  title = {Explainability as a non-functional requirement: challenges and recommendations},
  journal = {Requirements Engineering},
  year = {2020},
  volume = {25},
  pages = {493-514},
  doi = {10.1007/s00766-020-00333-1},
  language = {English},
  publisher = {Requirements Engineering}
}

@book{Wohlin2012,
  title     = "Experimentation in Software Engineering",
  author    = "C. Wohlin and P. Runeson and M. Höst and M. Ohlsson and B. Regnell and A. Wesslén",
  year      = 2012,
  doi = {10.1007/978-3-642-29044-20},
  publisher = "Springer",
}

@misc{Tsakalakis2022,
  author = {N. Tsakalakis and S. Stalla-Bourdillon and T. Huynh and L. Moreau},
  title = {A taxonomy of explanations to support Explainability-by-Design},
  year = {2022},
  month = {4},
  doi = {10.48550/arXiv.2206.04438},
  language = {English},
}

@article{Panichella2015,
  author = {S. Panichella and A. Sorbo and E. Guzman and C. Visaggio and G. Canfora and H. Gall},
  title = {How Can I Improve My App? Classifying User Reviews for Software Maintenance and Evolution},
  journal = {21st International Conference on Software Maintenance and Evolution},
  year = {2015},
  pages = {281 - 290},
  doi = {10.1109/ICSM.2015.7332474},
  language = {English},
  publisher = {IEEE}
}

@article{Pagano2013,
  author = {D. Pagano and W. Maalej},
  title = {User feedback in the appstore: An empirical study},
  journal = {21st International Requirements Engineering Conference},
  year = {2013},
  pages = {125 - 134},
  doi = {10.1109/RE.2013.6636712},
  language = {English},
  publisher = {IEEE}
}

@article{Landis1977,
  author = {J. Landis and G. Koch},
  title = {A Coefficient of Agreement for Nominal Scales},
  journal = {Biometrics},
  year = {1977},
  pages = {159-174},
  doi = {10.2307/2529310},
  language = {English},
  publisher = {International Biometric Society}
}

@misc{Drostedata2024,
  author       = {Droste, Jakob and
                  Deters, Hannah and
                  Obaidi, Martin and
                  Schneider, Kurt},
  title        = {{Supplementary Material - "Explanations in Everyday 
                   Software Systems: Towards a Taxonomy for
                   Explainability Needs" (RE'24)}},
  month        = mar,
  year         = 2024,
  publisher    = {Zenodo},
  doi          = {10.5281/zenodo.10871086},
  url          = {https://doi.org/10.5281/zenodo.10871086}
}

@online{StoreStatistiksEnglish2024,
  author = {Laura Ceci},
  title = {Number of apps available in leading app stores as of August 2024},
  year = 2024,
  url = {https://www.statista.com/statistics/276623/number-of-apps-available-in-leading-app-stores/},
  urldate = {2024-10-08}
}

@article{Kim2022,
  author = {E. Kim and H. Yoon and J. Lee and M. Kim},
  title = {Accurate and prompt answering framework based on customer reviews and question-answer pairs},
  journal = {Expert Systems with Applications},
  year = {2022},
  volume = {203},
  pages = {12},
  doi = {10.1016/j.eswa.2022.117405},
  language = {English},
  publisher = {Elsevier Ltd}
}

@article{Köhl2019,
  author = {M. Köhl and K. Baum and M. Langer and D. Oster and T. Speith and D. Bohlender},
  title = {Explainability as a non-functional requirement},
  journal = {27th International Requirements Engineering Conference},
  year = {2019},
  pages = {363-368},
  doi = {10.1109/RE.2019.00046},
  language = {English},
  publisher = {IEEE}
}

@article{Marcus2018,
  author = {G. Marcus},
  title = {Deep Learning: A Critical Appraisal},
  journal = {Computing Research Repository},
  year = {2018},
  pages = {27},
  doi = {10.48550/arXiv.1801.00631},
  language = {English},
  publisher = {Cornell University}
}

@article{Vermeulen2018,
  author = {B. Vermeulen and J. Kesselhut and A. Pyak and P. Saviotti},
  title = {The Impact of Automation on Employment: Just the Usual Structural Change?},
  journal = {Sustainability},
  year = {2018},
  pages = {27},
  doi = {10.3390/su10051661},
  language = {English},
  publisher = {Molecular Diversity Preservation International}
}

@article{Liu2023,
  author = {J. Liu and C. Liu and P. Zhou and R. Lv and K. Zhou and Y. Zhang},
  title = {Is ChatGPT a Good Recommender? A Preliminary Study},
  journal = {32nd ACM International Conference on Information and Knowledge Management},
  year = {2023},
  doi = {10.48550/arXiv.2304.10149},
  language = {English},
  publisher = {unpubliziert}
}

@article{Horstmann2023,
  author = {A.Horstmann and N. Krämer and C. Geminn and T. Bile and C.Weber and A. Manzeschke and L. Mavrina and S. Kopp and A. Artelt, B. Hammer},
  title = {Kann sich künstliche Intelligenz selbst erklären?},
  journal = {IMPACT},
  year = {2023},
  doi = {10.17185/duepublico/77378},
  language = {German},
  publisher = {Universität Duisburg-Essen}
}

@inproceedings{deters2024ExplainabilityAndUX,
author = {Deters, Hannah and Droste, Jakob and Hess, Anne and Kl\"{o}s, Verena and Schneider, Kurt and Speith, Timo and Vogelsang, Andreas},
title = {The X Factor: On the Relationship between User eXperience and eXplainability},
year = {2024},
isbn = {9798400709661},
publisher = {Association for Computing Machinery},
address = {New York, NY, USA},
url = {https://doi.org/10.1145/3679318.3685352},
doi = {10.1145/3679318.3685352},
booktitle = {Proceedings of the 13th Nordic Conference on Human-Computer Interaction},
articleno = {19},
numpages = {12},
location = {Uppsala, Sweden},
series = {NordiCHI '24}
}

\end{document}